# The Improved Job Scheduling Algorithm of Hadoop Platform

**Yingjie Guo[a], Linzhi Wu[b], Wei Yu[c], Bin Wu[d], Xiaotian Wang[e]**

[a,b,c,d,e] University of Chinese Academy of Sciences

100408, China

[b] Email: wulinzhi1001@163.com

[a,c,d,e] Email: {guoyingjie, yuwei, wubin, wangxiaotian}@ict.ac.cn

**[Abstract]** This paper discussed some job scheduling algorithms for Hadoop platform, and proposed a jobs scheduling optimization algorithm based on Bayes Classification viewing the shortcoming of those algorithms which are used. The proposed algorithm can be summarized as follows. In the scheduling algorithm based on Bayes Classification, the jobs in job queue will be classified into bad job and good job by Bayes Classification, when JobTracker gets task request, it will select a good job from job queue, and select tasks from good job to allocate JobTracker, then the execution result will feedback to the JobTracker. Therefore the scheduling algorithm based on Bayes Classification influence the job classification via learning the result of feedback with the JobTracker will select the most appropriate job to execute on TaskTracker every time. We need to consider the feature usage of job resource and the influence of TaskTracker resource on task execution, the former of which we call it job feature, for instance, the average usage rate of CPU and average usage rate of memory, the latter node feature, such as the usage rate of CPU and the size of idle physical memory, the two are called feature variables. Results show that it has a significant improvement in execution efficiency and stability of job scheduling.

**[Key Words]**: Hadoop; scheduling algorithm; improvement; Bayes

## 1. Introduction

Hadoop is an open source under the Apache fund account component, and is an open source implementation of Google graphs calculation model. It can easily develop and run large-scale data processing. Two of the most core part are HDFS (Hadoop Distributed File System) and MapReduce.[1]

**HDFS:** The Hadoop distributed file system (HDFS) to store large files with streaming data access patterns, to run with managers-workers mode, that is, there is a Name Node (managers) and multiple Data Nodes

(workers). Name Node manages the file system tree and the tree in all of the files and directories. Data Node is usually a Node in the cluster, a record of each file in each block of Data Node information.

**MapReduce:** MapReduce work process is divided into two phases: the Map and Reduce phase. A Map function, which is used to put a set of keys for mapping into a new set of key-value pairs. And it points to the Reduce function. MapReduce has four parts: the framework of homework submission and initialization, task allocation, task execution and completion of the homework. [2]

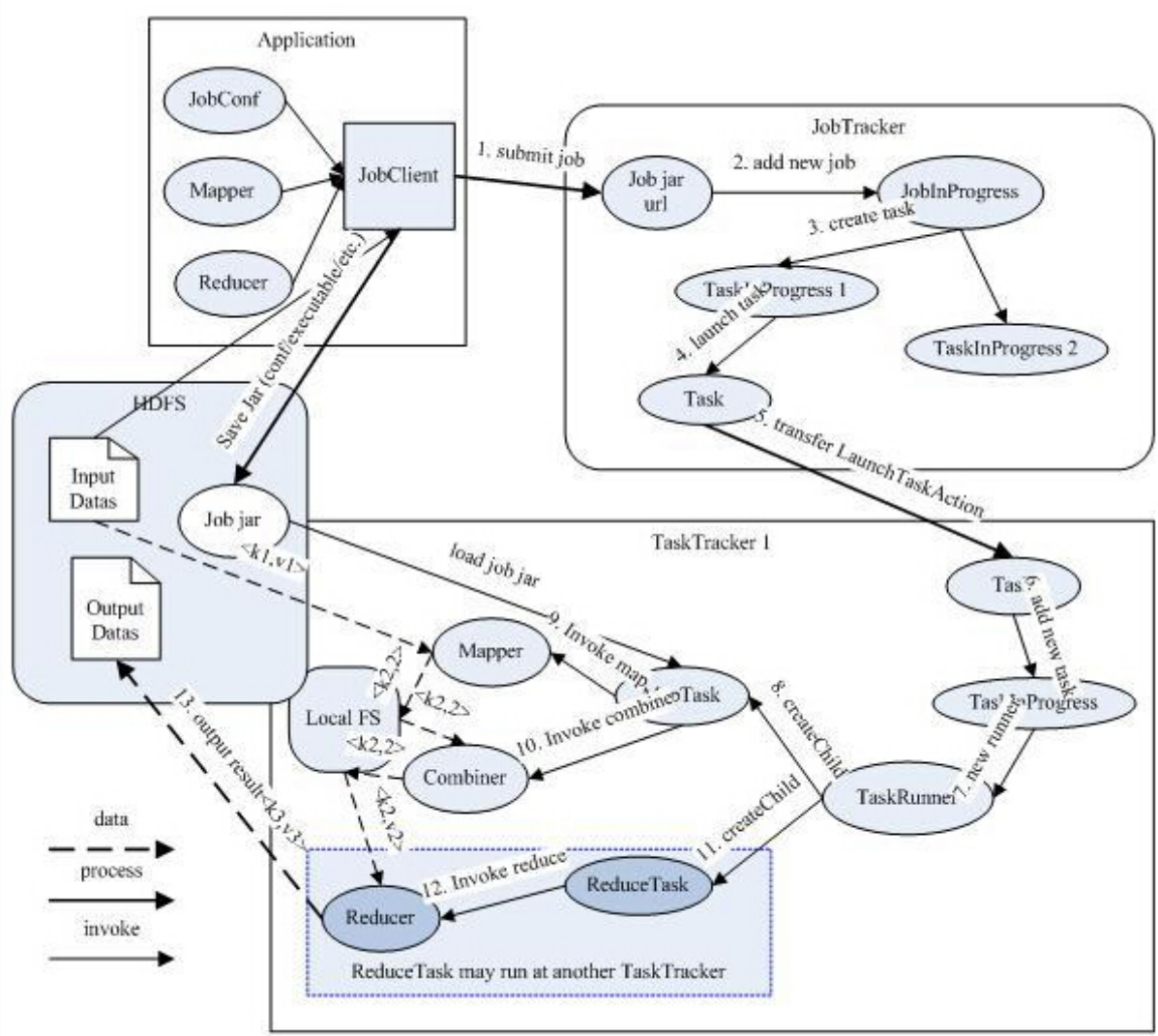

**Figure 1. Hadoop MapReduce Framework Architecture**

Firstly user program (Job Client) submits a job, and then the job of the information will be sent to the job Tracker. Job Tracker is the center of the Map - reduce framework, which needs to communicate with the cluster machine timing (heartbeat), and need to manage what program should be run on which machines, to manage job failed, restart operation. TaskTracker is a part of each machine in MapReduce. It is designed to surveillance resources of their machines. TaskTracker monitoring tasks run of the current state of the machine. TaskTracker needs sends the information through the heartbeat JobTracker. JobTracker will collect these information to assign new job submitted a run on which machines.

## 2. Hadoop YARN

### 2.1. The background of Hadoop YARN

The initial MapReduce architecture is simple and clear. In the first few years of launch, it also got lots of successful cases, won the industry wide support and affirmation. But as a distributed system of the cluster scale and its workload growth, the original framework of problem gradually emerges, the main problem is as follows:

- The JobTracker is the point of focus on the Map-reduce, there is a single point of failure;

- The JobTracker completed the most of task, causing too much resource consumption. When the map - reduce job very much, it can cause a large memory overhead, also increase the JobTracker fail risk, it is also the industry generally summed up the old Hadoop map - reduce the cap on the host only can support 4000 nodes;

- In TaskTracker, framework doesn't take into account the CPU/memory usage. If two large memory consumption of the task to be scheduled one, it is easy to appear OOM;

- The JobTracker is the point of focus on the Map-reduce, there is a single point of failure;

- In TaskTracker, the resources are divided into the map task slot and reduce task slot. If only map task or only the reduce task in the system, it will cause the waste of resources.

### 2.2. The framework of Hadoop YARN

YARN is the resource management system in the Hadoop 2.0. It splits the JobTracker of MRv1 into two independent service: a global resource manager named ResourceManager and ApplicationMaster of each application. The ResourceManager is responsible for the resource management and allocation of the whole system, while ApplicationMaster responsible for the management of a single application. [3]

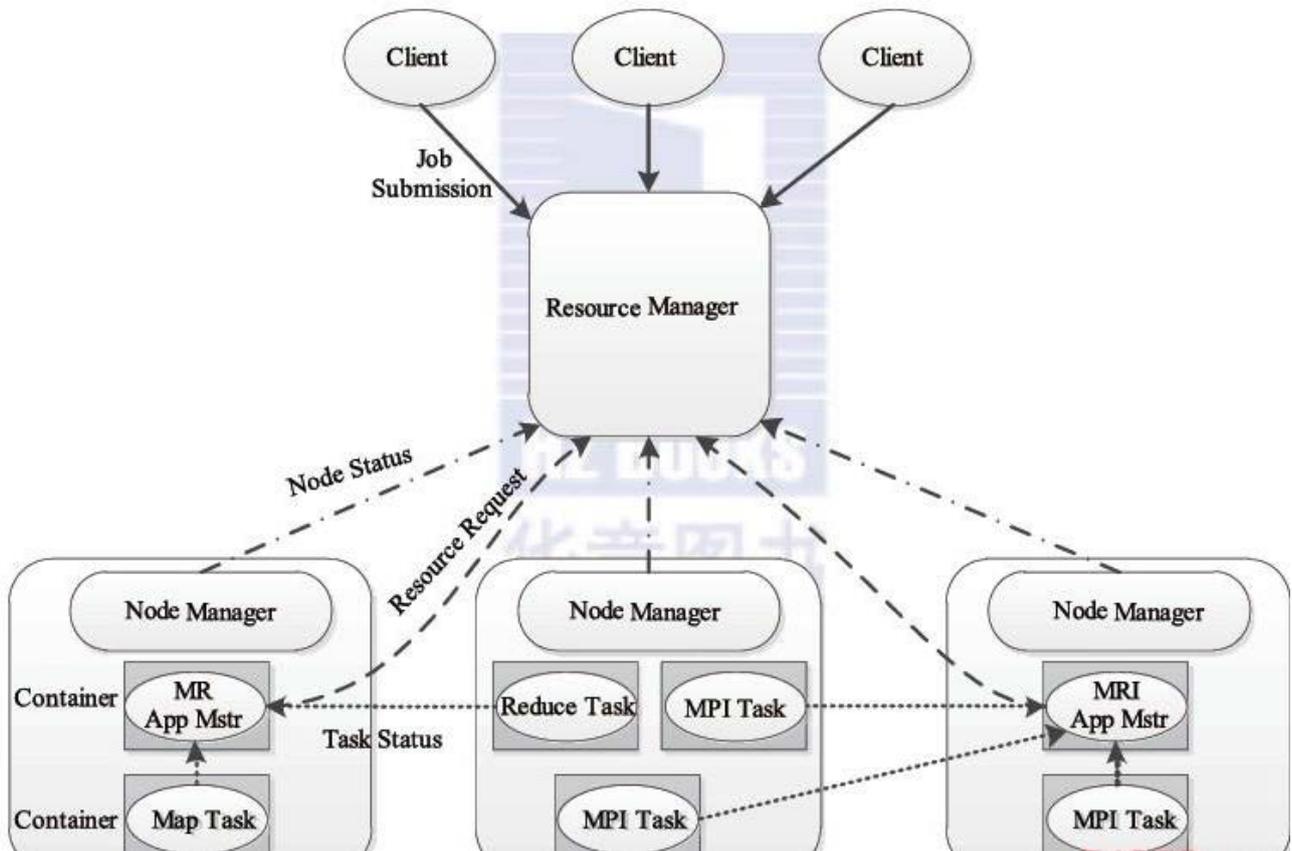

**Figure 2. YARN Framework Architecture**

YARN is still the Master/Slave structure. In the resource management framework, the ResourceManager is Master, NodeManager is a Slave, and the ResourceManager is responsible for all the resources on the NodeManager for unified management and scheduling. YARN is mainly composed of the ResourceManager, NodeManager, ApplicationMaster and several Container components.

•ResourceManager (RM): RM is a global resource manager, is responsible for the resource management and allocation of the whole system. It is mainly made up of two components: the Scheduler (Scheduler) and the application Manager (Applications Manager, ASM);

•ApplicationMaster (AM): Each application contains 1 AM. There are the main features: Negotiate with RM scheduler for resources, Tasks within the task assigned to further, Communicate with NM to start/stop the task, the Monitor all tasks running state;

•NodeManager (NM): NM is on each node of resources and task manager. On the one hand, it will report regularly to the RM this node on the resource usage and the running state of every Container. On the other hand, it receives and deal with the Container from AM start/stop and other requests;

•Container: Container is resource abstraction of the YARN. It encapsulates the multi-dimensional resources on a node, such as memory, CPU, disk, network and so on.

## 2.3. The process of Hadoop YARN

When the user submit an application to the YARN, YARN will run the application with the following steps:

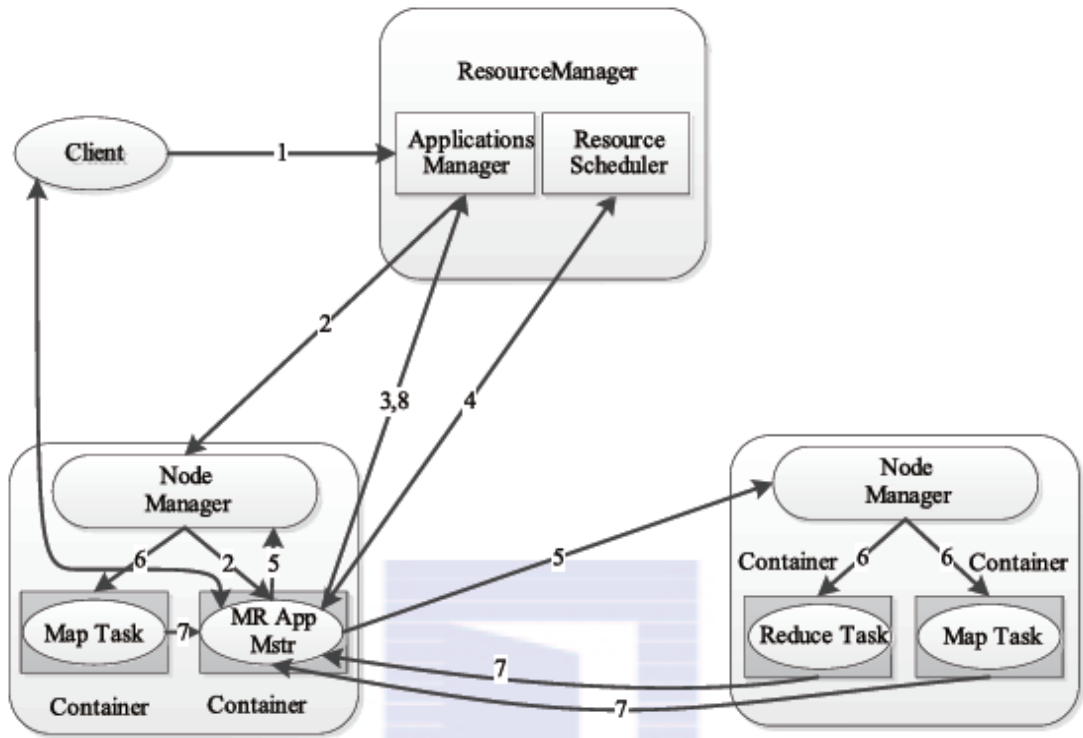

**Figure 3. YARN working principles**

• User submit the application to the YARN, including ApplicationMaster program, startup ApplicationMaster commands and user program and so on;

• The ResourceManager distribute the first Container for this application. It communicates with the Node - Manager, required to start the application in the Container of ApplicationMaster;

• ApplicationMaster register with the ResourceManager firstly. The user can view application running state directly through the ResourceManager;

• ApplicationMaster adopt the way of polling by RPC protocol applied to the ResourceManager for and the resources;

• After ApplicationMaster get the resource, it will communicate with the corresponding NodeManager and start the task.

• After NodeManager set up running environment, task start command is written into a script and start the task;

- Through the RPC protocol, each task report their own status and progress to ApplicationMaster. ApplicationMaster can know the running status of each task;

- At the end of an application, ApplicationMaster will log out and close itself.

## 3. Introduce Of Hadoop Scheduler

In the platform of Hadoop, the schedule of homework is to assign the proper tasks to proper servers. There are two steps to go. Firstly, you should select the homework, then in the homework you should choose the right task. Just the same as the work of assigning, assigning the tasks is a complex job. The bad assigning of tasks results in the increments of mount of network. Some task servers may overload and be lower effectively. And the assigning tasks is complex thing, never can you solve the problem by one way. According the different backgrounds, you may need the different methods to feed the needs. In the platform of Hadoop, there are three ways of homework scheduling. They are FIFO, fair scheduling method, and the computing scheduling method.

### 3.1. FIFO

Hadoop chooses the method of FIFO as default. [4] It chooses the homework to execute by the priority of the homework and the turns of arriving. First come, and first go. FIFO is quite simple. In the platform of Hadoop there is only one homework queue. And the submitted homework will be queued by time. The new coming task will be inserted in the tail of the task queue. When the task is finished, it always pick the head of the queue. The advantage of this scheduling is simple and easy to realize. But the disadvantage of this scheduling is obviously. It sees all the homework as the same without regarding of the time of homework. And this kind of scheduling is bad to small homework.

Advantage**:** easy to realized, and the process of scheduling is fast.

Disadvantage: usage of resource is low.

### 3.2. Fair scheduling strategy

This method configures the task pool in the system, and a task pool can execute a task. These tasks are the result of a big homework divided. When one user submit many homework, every task is assigned to some task pool to execute the tasks. If comparing the homework scheduling of homework to the scheduling of system, the first FIFO is like some bat tasks. Every moment there is only one task executed .And the fair scheduling task just likes bat dealing system. It realized the task of executing many homework at the moment. Because the Linux is grouped, if some users submit some homework at the same time. At this strategy, it will assign every user a homework pool. Then it will set the amount of minims sharing pools.

The minus task .To make sure .when the current empty task pool is released, it will be assigned to this homework pool. First understand what it meant a minimum, the minimum means as long as the job pool needs, the scheduler should be able to meet this requirement to ensure that the minimum number of jobs task slot pool, but how to ensure that when it takes on the task of empty slots, one way is to allocate a certain fixed number of slots to the job pool is not fixed, the amount is at least the minimum task slot values, so long as the time required for the job pool allocated to it the line, but this pool is not used in this job so multitasking time slot wasteful, this strategy is actually doing when demand does not meet the minimum job pool task slot number, nominally own the remaining tasks slots will be distributed to other needy jobs pool, when the need to apply for a job pool task slot if the system is not, at this time not to seize other people's (do not know who robbed ah), as long as the current release of an empty slot task will be assigned to the immediately job pool.

In the homework pool of the user, how to assign the homework to the pool .It can choose the FIFO. So the scheduling of strategy is divided into two levels. The first level: every user is assigned to a homework pool in the moment of assignment of some users. The second level: every user can use different scheduling strategy.

Advantage:

It supports the scheduling of divided homework. The different type of tasks will get the different assignment of resources. It will promote the quality of service, and adjust the amount of homework executed the same time. And it will use resources deeply.

Disadvantage:

It ignores the node of the balance states, and it will result in unbalance.

3.3. The computing scheduling

The computing scheduling is a little like the fair scheduling. The fair scheduling assign the task pools by the unit of homework. But the computing scheduling assigns the tasks by the unit of task tracker. This kind of scheduling configures some queues. When a queue has a free TaskTracker, the scheduling will be assigned to other queue. When there is having free TaskTracker, it may face the solution some queues get no the enough amount TaskTracker. Free TaskTracker will be assigned to the hungriest queue. How to measure the hungry level? It can be judged by the result of the amount of executing tasks and the computing resources. The lower, the more hungry it is. Capacity scheduling strategy is organized as a cohort of homework, so a user operation may in multiple queue, if the user does not do certain restrictions, is likely to occur serious phenomenon of unfair between multiple users. So when the new job is selected, it is also required to consider whether the user of the job is more than the limit of resources, if more than, the job will not be selected. For the same queue, this strategy uses a priority based FIFO policy, but will not preemption.

Advantage:

It supports the scheduling of divided homework. It will promote the quality of service, and adjust the amount of homework executed the same time. And it will use resources deeply.

Disadvantage:

User need known a large amount of system information, then user can set the setting and choose the queue.

## 4. Scheduling Algorithm Based On Naive Bayes

4.1. Background

A job is divided into multiple tasks and job scheduling implements the function that distribute the tasks of a job to a TaskTracker to be completed, the point of job scheduling is how to choose an appropriate job distribute the tasks of it to appropriate TaskTracker to execute. In all of the available scheduling algorithms, if we want to implement these functions, we need to preset the program. Firstly, the administrator needs to set the maximum number of tasks which can run simultaneously, if the setting of maximum tasks is too large, the resources of every task acquired will be too small, the running of tasks will result in overload and inflect the completion of the tasks, and if the setting of tasks is too small, it will lead to the tasks obtain surplus resources which will influence the using efficiency of system resources. However, only when the administrator has enough knowledge of the MapReduce job and TaskTracker, can the setting be appropriate, which is really too hard, especially in the situation where there exists lots of MapReduce Jobs and facing with the problem that a Hadoop cluster possess many TaskTracker. [5] Besides, the setting of resources allocation of Job Pool and Job Queue are needed for fair-based algorithm and computing capacity based algorithm. Therefore, since so many discrepancies exist, preset does not match with the current demand. We really need an algorithm which adjusts the selection of Job and task allocation by constantly monitoring or learning the actual resource using during the execution of the task to achieve each task assigned to the task assigned to the most appropriate to complete the task running on TaskTracker. On this basis that we put forward the scheduling Algorithm Based on Bayes Classification.

4.2. Algorithm Introduction

In the scheduling algorithm based on Bayes Classification, the jobs in job queue will be classified into bad job and good job by Bayes Classification, when JobTracker gets task request, it will select a good job from job queue, and select tasks from good job to allocate JobTracker, then the execution result will feedback to the JobTracker. Therefore the scheduling algorithm based on Bayes Classification influence the job classification via learning the result of feedback with the JobTracker will select the most appropriate job to execute on TaskTracker every time. [6] We need to consider the feature usage of job resource and the influence of TaskTracker resource on task execution, the former of which we call it job feature, for instance, the average usage rate of CPU and average usage rate of memory, the latter node feature, such as the usage rate of CPU and the size of idle physical memory, the two are called feature variables.

These two types of feature variable will be introduced in details below.

**Job feature:** this kind of feature variable mainly describes the resource usage situation of job. The value of the variable can be set when the user commits job or obtained via analysis the history information of job execution. We will use the former one in this context. The variable values are set from 10 to 1, and 10 is the maximum value which represents the utmost using of resources, 1 corresponding to the minimum value which represents the min usage of resources. The feature variables average CPU usage rate of job, average network usage rate, and average usage rate of IO and average memory usage rate will be adopted.

**Node feature:** this kind of feature variable represents the computation resource state and quality on a TaskTracker computing node. And the variable can be divided into static feature variable of node and dynamic feature variable of node. The static feature variable refers to such feature variables which stay static or are constants. While the dynamic feature variable refers to those node properties which change frequently along with time.

As for job feature, the higher the value, the more it uses computation resources, therefore, the possibility of overloading will be higher. While for node feature, the lower the value, the lower usability of computation resources, and the possibility of overloading will be higher. [7] This also means that, for any feature variables, there is a critical value which above or below will make the node of job overload. A hyper-plane is composed by all of these critical values, thus we can use a linear classifier to divide jobs with the hyper-plane as decision-making plane and because the feature variables we use can fit the conditional independence assumption, so we choose the naive Bayes classifier as our classifier.

We divide jobs into two categories, we call the one which will not make TaskTracker overload good job. Relatively, we call the one which will make TaskTracker overload bad job. When classifying, we will calculate the value $P(a_i=good|J_1,J_2,....,J_n)$ and $P(a_i=bad|J_1,J_2,....,J_n)$ separately, where $P(a_i=good|J_1,J_2,....,J_n)$ represents the probability that a job i is a good job for a value of feature variable $J_1, J_2, ..., J_n$, and $P(a_i=bad|J_1, J_2, ...., J_n)$ represents the probability that a job i is a bad job for a value of feature variable $J_1, J_2, ..., J_n$, whether a job is good or bad all depends to the probability of bad job and good job.

For $P(a_i=good|J_1, J_2, ...., J_n)$, we can get the formula just as below when we use Bayes Theory.

$$P(a_i=good|J_1, J_2, ...., J_n) = \frac{P(J_1, J_2, ..., J_N | a_i = good) P(a_i = good)}{P(J_1, J_2, ..., J_n)}$$

And according to Bayes conditional independent assumption, we can get the following formula:

$$P(J_1, J_2, ..., J_n | a_i = good) = \prod_{j=1}^{n} P(J_j | a_i = good)$$

Besides, the grave part of the formula has nothing to do to job, so its value can be neglect wile calculating. Therefore, we can get the following formula finally:

$$P(J_1, J_2, ..., J_n | a_i = good) = P(a_i = good) \prod_{j=1}^{n} P(J_j | a_i = good)$$

Similarly,

$$P(J_1, J_2, ..., J_n | a_i = bad) = P(a_i = bad) \prod_{j=1}^{n} P(J_j | a_i = bad)$$

$P(a_i=good), P(a_i=bad), P(J_j|a_i=good), P(J_j|a_i=bad) (j=1,...,n)$ are all Prior Probability, their values are updated through the execution of every task allocated to a TaskTracker, an then calculate $P(a_i=good|J_1, J_2, ...J_n)$ and $P(a_i=bad|J_1, J_2, ...J_n)$ by their values.

We call the rule which determine whether the execution of task allocation leads to the TaskTracker which it execute on overload overloading rule. The judgment basis of the overloading rule is the resource using information of the TaskTracker, and the criteria for judging are based on the specific demands of user to develop, for example the most jobs are CPU intensive ones, then the usage rate of CPU can used to be the standard to determine whether the job overloads. In addition we are not limited to just one judgment standard but synthesis multiple conditions for judging. For example we can synthesis the usage condition of CPU, memory, network and so on as overload standards.

The judgment result of overloading rule will feed back to classifier to update the values of $P(a_i=good), P(a_i=bad), P(J_j|a_i=good), P(J_j|a_i=bad) (j=1,...,n)$, so we can calculate $P(a_i=good|J_1, J_2, ...J_n)$ and $P(a_i=bad|J_1, J_2, ...J_n)$, and determine whether the job is good or bad under present condition of feature variables according to the result of calculation. Then, we choose the good one to allocate. If there are many jobs are classified as good jobs, then we will choose the one which makes the following formula has maximum.

$$E.U.(i) = U(i)P(a_i=good|J_1,J_2,...,J_n)$$

Among them, E.U.(i) is expected utility, U(i) is the utility function which related to job i. $a_i$ represents job i. Once a job is selected, we need to select the required data in the job to schedule the tasks on the TaskTracker firstly. If there does not exist such kind of tasks, we will select the tasks whose data are not local to schedule. Here, we import utility function to set the prior level of jobs and implements some scheduling strategies. Without utility function, the scheduler will always select the jobs which can provide maximum system availability among those are classified good ones. Every time allocate tasks to a TaskTracker, we will observe the effect of the last task allocation via the information of the Task Tracker's next hop. If we find the TaskTracker overload, we need to train classifier to avoid similar mistake. Specifically, the feedback of the judging results of the overloading rule is given to the classifier to update the value of $P(a_i=good), P(a_i=bad), P(J_j|a_i=good), P(J_j|a_i=bad)$, so as to affect the job classification and job selection.

4.3. Algorithm analysis

One of the most important characters of Job scheduling algorithm based on Bayes classification is that the administrator can adjust task allocation policy through learning the feedback result of tasks lies to node constantly without knowledge of the resource using feature of MapReduce job and resource of TaskTracker on the cluster to improve the correct rate of task allocation and then provide the most system availability.

## 5. Conclusion

This paper has discussed the background, platform architecture and core part of Hadoop platform in details, and then we put forward the Job scheduling algorithm Based on Bayes Classification viewing the shortcoming of those algorithms which are used. The Existing algorithms require administrator has enough knowledge about the resources using feature of jobs running on MapReduce and TaskTracker resources on cluster, in addition, the administrator need to negotiate with users of cluster of their share to ensure the correct operation of jobs. However, when facing with a big cluster possessing huge number of users, it will enhance the burden of administrator and result in unnecessary error. The Job Scheduling algorithm Based On Bayes Classification promoted in this paper can avoid this problem well, it needn't administrator has enough knowledge about the resources using feature of jobs running on MapReduce and TaskTracker resources on cluster, it will ensure the job complete correctly with the learning the feedback result of every task allocation decision to cluster resources to affect or adjust later allocation strategy constantly, this can reduce administrator's burden, improve management efficiency and reduce the possibility of human error obviously.